\newcommand{\beq}{\begin{eqnarray}}
\newcommand{\eeq}{\end{eqnarray}}

\documentclass[prl,twocolumn]{revtex4}
\usepackage{graphicx,amssymb}

\begin{document}


{\bf Comment on ``Temporal scaling at Feigenbaum point and nonextensive thermodynamics" by P. Grassberger}

Recently, P. Grassberger \cite{grass} focused on some relevant aspects of nonextensive statistical mechanics, based on the entropy $S_q\equiv k\frac{1-\sum_ip_i^q}{q-1}$ \cite{history}. His Letter contains, side by side, correct results (most of them long known), and severely incorrect statements. The whole appears to reflect a somewhat abridged understanding of the theory that the author criticizes.  The present Comment addresses some of the latter.

{\bf 1-} To start with, it is by now widely known that the $q \ne 1$ theory applies to macroscopic states that exhibit some kind of (quasi-) stationarity or metastability, and by no means applies to thermal equilibrium, which is definitively well described by standard statistical mechanics. This point and others are clarified in \cite{reply}, skipped in Ref. 6 of \cite{grass}.   

{\bf 2 -} For the Shannon and related entropic forms, it is reserved in \cite{grass} expressions such as ``the only rationally justifiable ansatz" and ``the only consistent probabilistic measure of information".  Regretfully, on one hand this ignores  the words of Shannon himself ({\it ``This theorem and the assumptions required for its proof, are in no way necessary for the present theory. It is given chiefly to lend a certain plausibility to some of our later definitions. The real justification of these definitions, however, will reside in their implications."}\cite{shannon}). On the other hand, it also ignores relevant theorems and analytical properties that $q$-generalize those which currently characterize the Shannon entropy $S_1$. Let us mention a few of them. 

(i) The Santos \cite{santos} theorem, which $q$-generalizes the Shannon theorem: If an entropy $S$ is a continuous function of $\{p_i\}$, {\it and} monotonically increases with $W$ in the case of equal probabilities (i.e., $p_i=1/W$), {\it and}  satisfies $S(A+B)/k=S(A)/k+S(B)/k+(1-q)S(A)S(B)/k^2$ for all independent subsystems $A$ and $B$ ($k$ is a conventional positive constant), {\it and} satisfies the grouping property $S(\{p_i\})=S(p_L,p_M)+p_L^q \,S(\{ p_i/p_L\})+p_M^q \, S(\{ p_i/p_M\})$ ($p_L+p_M=1$; $\{p_i/p_L\}$ and $\{p_i/p_M\}$ are the conditional probabilities), {\it then and only then} $S=S_q$. 

(ii) The Abe \cite{abe} theorem, which $q$-generalizes the Khinchin theorem: If an entropy $S$ is a continuous function of $\{p_i\}$, {\it and} monotonically increases with $W$ in the case of equal probabilities (i.e., $p_i=1/W$), {\it and} $S(p_1,p_2,...,p_W,0)=S(p_1,p_2,...,p_W)$, {\it and}  satisfies $S(A+B)/k=S(A)/k+S(B|A)/k+(1-q)S(A)S(B|A)/k^2$ for arbitrary subsystems $A$ and $B$ ($k$ is a conventional positive constant; $S(A|B)$ is the conditional entropy), {\it then and only then} $S=S_q$. 

(iii) $S_q(\{p_i\})$ is concave for all $q>0$ (Ref. 2 of \cite{grass}). 

(iv) $S_q(\{p_i\})$ is Lesche-stable (also referred to as {\it experimentally robust}) \cite{lesche}. 

(v) $S_q$ can be extensive: $q=1$ for independent (or almost independent) subsystems, and $q \ne 1$ for some (strictly or asymptotically) scale-invariant systems \cite{TGS} (Moreover, this point is related to a possible generalization of the standard Central Limit Theorem \cite{CLT}).  

(vi) $S_q(t)$ leads, for a special value of $q$ and appropriate limits (including $t \to\infty$), a {\it finite} entropy production per unit time, where  $S_q(t)$ denotes an ensemble-based entropy (defined in detail below, in \cite{GT}, and elsewhere). This entropy production is a concept close (though {\it different}) to the so called Kolmogorov-Sinai entropy. Recent illustrations include those analytically discussed in Ref. 36 of \cite{grass} and in \cite{casati}.

It is quite remarkable the fact that $S_q$ and $S_1$ share so many basic properties. It is worthy to mention, for instance, that Renyi entropy -- certainly useful in the geometric characterization of chaotic dynamical systyems -- violates all the above properties (iii)-(vi) (to be more precise, it satisfies (v) for {\it all} $q$ if the subsystems are {\it (quasi-) independent}, but it violates it for generic globally correlated subsystems). 

{\bf 3 -} We read in the Abstract of \cite{grass} that ``there is no generalized Pesin identity for this system". Let us revisit this point. On one hand, at the edge of chaos of  one-dimensional unimodal dissipative maps such as the $z$-logistic one, it has been conjectured in Ref. 8 of \cite{grass}, and analytically proved in Ref. 19 of \cite{grass} (and proved once again in \cite{grass}, Eq. (3)), that the (upper bound of the) sensitivity to initial conditions $\xi \equiv \lim_{\Delta x(0) \to 0} \Delta x(t)/\Delta x(0)$  is given by $\xi=e_{q_{sen}}^{\lambda_{q_{sen}} t}$, where $\lambda_{q_{sen}}$ generalizes the usual Lyapunov exponent (here recovered as the $q_{sen}=1$ particular instance); {\it sen} stands for {\it sensitivity}. (Notice, by the way, that $\xi$ is the exact solution of $d\xi/dt=\lambda_q  \,\xi^q$. Many more details can be found in Ref. 36 of \cite{grass} and in \cite{robledomori}). On the other hand, we can partition the phase space $x \in [-1,1]$ of the system $x_{t+1}=1-a |x_t|^z$ in $W$ little cells (denoted by $i=1,2,...,W$). Within one of these cells we can put $M$ initial conditions. We can then run the map for all these points and define the set of probabilities $p_i(t) \equiv M_i(t)/M$ (with $\sum_{i=1}^W M_i(t)=M$). With these probabilities we can calculate $S_q(t)$, and focus on the {\it supremum} of such function over all partitions and initial cells. It follows that an unique value of $q$ exists, {\it precisely} $q_{sen}$, such that $K_q \equiv \lim_{t \to\infty} \lim_{W \to \infty} \lim_{M \to \infty} S_q(t)/t$ is {\it finite} ($K_q$ vanishes for $q>q_{sen}$, and diverges for $q<q_{sen}$). It has been conjectured in Ref. 8 of \cite{grass}, and analytically proved in Ref. 36 of \cite{grass}, that $K_{q_{sen}}=\lambda_{q_{sen}}$. The particular case $q_{sen}=1$ yields $K_1=\lambda_1$. This relation is {\it not} the same as Pesin identity, which uses, not the present $K_1(t)$, but rather the Kolmogorov-Sinai entropy rate. It is, however, totally similar, which makes that it is sometimes referred in the literature as Pesin-like, or simply Pesin identity.

{\bf 4 -} We also read in \cite{grass} that ``In the first papers [...], it was supposed but never substantiated that the parameter $q$ of NET can be obtained, also for the Feigenbaum map, by some maximum entropy principle." This is a strange statement indeed. From the very first steps of the nonextensive theory, the entropic index $q$ was assumed to have some special value (obtained in fact {\it a priori} from microscopic dynamics, in agreement with Einstein's and Cohen's philosophies \cite{EinsteinCohen}). Under appropriate circumstances, extremization of $S_q$ is to be seeked for characterizing stationary or stationary-like states (in close correspondence with thermal equilibrium of Hamiltonian systems, which are known to extremize $S_1$). Why the author of \cite{grass} appears to suggest that $q$ itself  would be determined through some maximum entropy principle remains as some sort of mystery.     
   
There are several other points in \cite{grass} that would deserve comments were it not space limitations. However, we have illustrated that, by its poor knowledge of the relevant literature, \cite{grass} can be regretfully misleading for the nonspecialized readers.

\noindent
{\it Constantino Tsallis},  
Santa Fe Institute,
1399 Hyde Park Road,
Santa Fe, New Mexico 87501,  USA, {\it and}
Centro Brasileiro de Pesquisas F\'\i sicas,\
Rua Xavier Sigaud 150, 
22290-180 Rio de Janeiro-RJ, Brazil (tsallis@santafe.edu)

\end{document}